# Radio Tomography for Roadside Surveillance

Christopher R. Anderson, *Senior Member, IEEE*, Richard K. Martin, *Member, IEEE*, T. Owens Walker, *Member, IEEE*, and Ryan W. Thomas, *Senior Member, IEEE*

*Abstract*—Radio tomographic imaging (RTI) has recently been proposed for tracking object location via radio waves without requiring the objects to transmit or receive radio signals. The position is extracted by inferring which voxels are obstructing a subset of radio links in a dense wireless sensor network. This paper proposes a variety of modeling and algorithmic improvements to RTI for the scenario of roadside surveillance. These include the use of a more physically motivated weight matrix, a method for mitigating negative (aphysical) data due to noisy observations, and a method for combining frames of a moving vehicle into a single image. The proposed approaches are used to show improvement in both imaging (useful for human-in-the-loop target recognition) and automatic target recognition in a measured data set.

*Index Terms*—Radio tomography, received signal strength, sensor network, surveillance.

## I. Introduction

UBIQUITOUS position awareness is a recurrent theme in many disparate areas of signal processing. One may wish to know one's own position (i.e., navigation), the position of resources, or the position of other entities. In image-based positioning, the object must not be obstructed and illumination conditions must be good. Radio Frequency (RF) based techniques are a popular alternative, since they operate regardless of illumination or field of view. If the object to be tracked has a radio transmitter, then the problem is called source localization [1]–[3]. However, in many applications, the object in question may not wish to carry a transmitter (e.g., patients in a nursing home, criminals, or military forces) or if they do, the emitting radio device may be deliberately non-standard (e.g., in law enforcement and military applications). To this end, Wilson and Patwari recently proposed Radio Tomographic Imaging (RTI) as a means of locating an object via radio waves even when the object in question does not have a measurable radio signal [4]. Although RADAR systems provide an ability to locate or track a non-emitting object, they have limited use in small-scale environments. In these environments, an object could be obscured by a significant amount of clutter or located behind walls, buildings, or dense vegetation.

In RTI, each sensor in a Wireless Sensor Network (WSN) repeatedly sends packets to all other sensors. If an object is physically obstructing a given link, then the link's Received Signal Strength (RSS) will drop relative to a pre-calibrated value. By observing link attenuations, it is possible to determine which voxels are occupied by obstructions, as well as how dense those obstructions are (relative to RF propagation).

Worldwide increases in human and drug trafficking, terrorist activities, and illegal border crossings have created a need for a real-time monitoring system that can identify and classify potential threats from a safe stand-off distance. Current security schemes to mitigate these threats use manpower-intensive ground patrols or checkpoints, as well as expensive aerial surveillance from manned or unmanned aircraft. These security schemes are extremely resource-intensive, and, as a result, may leave some vulnerable points exposed. In particular, the ability to ensure safe access to government-controlled sites, border crossings, and heavily traveled roadways in harsh environmental conditions using a minimum of resources remains a primary security concern. The combination of a WSN with RTI offers a low-cost potential to remotely monitor a roadway for a minimal investment of resources [5], [6]. Such a system would not only provide intrusion detection, but affords the capability to implement a "smart environment" [7]–[10]. The smart environment would combine the WSN capabilities of long-term remote monitoring and data fusion from other sensor technologies with the RTI-enabled benefits of RF passive object detection, tracking, and classification.

There have been relatively few papers in the literature on RTI or RTI enhanced WSNs. The series of papers by Wilson and Patwari, notably [4], are the seminal papers in this area, and they define the baseline model, reconstruction algorithms, and performance analysis. Additionally, enhanced processing techniques developed by the authors in [11] allow them to locate and track a user even when the physical environment changes over the course of a day. Kanso and Rabbat [12], [13], discuss distributed algorithms and compressive sensing approaches to RTI, and their physical model is slightly different than that in [4]. The work in [14] and [15] was done independently from [4] and uses the term Device free Passive (DfP) localization rather than RTI, but the idea is similar. The authors measure the mean and variance of the RSS on all links of a sensor network, and compare current conditions to a pre-calibrated database. The distinction

Manuscript received January 08, 2013; revised June 04, 2013; accepted October 02, 2013. Date of publication October 22, 2013; date of current version January 15, 2014. This work was supported in part by the Office of Naval Research and the Air Force Office of Scientific Research. The views expressed in this paper are those of the authors, and do not reflect the official policy or position of the United States Air Force, Navy, Department of Defense, or the U.S. Government. This document has been approved for public release; distribution unlimited. The guest editor coordinating the review of this manuscript and approving it for publication was Prof. Michael Rabbat.

C. R. Anderson and T. O. Walker are with the Wireless Measurements Group, U.S. Naval Academy, Annapolis, MD 21402 USA (e-mail: canderso@usna.edu; owalker@usna.edu).

R. K. Martin and R. W. Thomas are with the Department of Electrical and Computer Engineering, The Air Force Institute of Technology (AFIT), Wright-Patterson AFB, OH 45433 USA (e-mail: richard.martin@afit.edu; ryan.thomas@us.af.mil).

Color versions of one or more of the figures in this paper are available online at http://ieeexplore.ieee.org.

Digital Object Identifier 10.1109/JSTSP.2013.2286774

is that RTI uses a physical model whereas DfP uses a statistical model only and looks for aberrations. Follow-on work by the authors in [16] investigated using DfP to detect the presence and estimate the speed of vehicles based on a multi-class support vector machine applied to received signal strength. In [17], the goal was obstacle mapping rather than change detection. As such, there was no calibration period, and their physical model was much more detailed than in other RTI work. In [18], first RTI is used to estimate attenuating objects in order to improve the fading model for active RSS geolocation techniques. It is then shown that the RSS geolocation estimates are more accurate, since the attenuation model is more accurate. Zhang, *et al.* developed a DfP system (which they termed "Transceiver-free") based on a geometric technique that used a dynamic cluster-based probabilistic algorithm to solve the localization problem [19], [20]. In [21] the authors formulated the DfP problem as sparse signal reconstruction, and propose an algorithm based on compressive sensing techniques to estimate the location of an object in the network. Finally, [22] uses a motion model and a Kalman filter to improve RTI performance for dynamic scenes.

This work focuses on vehicle identification and tracking using an RTI network. Compared to RTI tracking of humans, this is much more challenging, since the vehicles can move significantly between scans of the network. Vehicles are also much larger than humans, and obtaining enough voxels to identify a vehicle's class requires that the network monitor a much larger area than what has been reported in the literature for tracking humans. Finally, the nodes cannot completely surround the vehicle, since they cannot be placed in the road; so the quality of the information in the cross-road dimension tends to be poor.

The outline of this paper is as follows. First, in Section II we review the RTI models used in the literature, and put them into a common framework. Next, Section III discusses design and implementation issues for setting up a 3D roadside surveillance network. Then Section IV proposes new solution techniques for the obstruction estimation problem. Finally, Section V illustrates the results using measured field data. The contributions of this paper include a variety of modeling and algorithmic improvements, namely a more physically motivated weight matrix, a method for incorporating more realistic priors, a method for mitigating negative (aphysical) data due to noisy observations, and a method for combining frames of a moving vehicle into a single image; as well as a simple vehicle class identification algorithm. Moreover, the proposed approaches are used to show improvement in both imaging (useful for human-in-the-loop target recognition) and automatic target recognition in a measured data set.

## II. REVIEW OF RTI

Consider a WSN consisting of $K$ sensors. The convex hull of the WSN is divided into $N$ voxels, which for simplicity are cuboids of size $\delta_x \times \delta_y \times \delta_z$ m$^3$. (In 2D a hexagonal tessellation of voxels may be worth consideration since it only requires 86% as many samples as a square tessellation to meet the Nyquist spatial sampling criterion [23], though that does not extend to 3D since a cube is the only Platonic solid that tessellates a 3D space.) For generality, the 3D term "voxel" will be used in place of the 2D term "pixel" even when examples are given in 2D. The WSN can be fully connected with $M = K(K-1)/2$ unique links; or the WSN can be split in two across a road as in Fig. 1, leading to $M = (K/2)^2$ links. An example of an RTI system is shown in Fig. 1, and videos are available at [24].

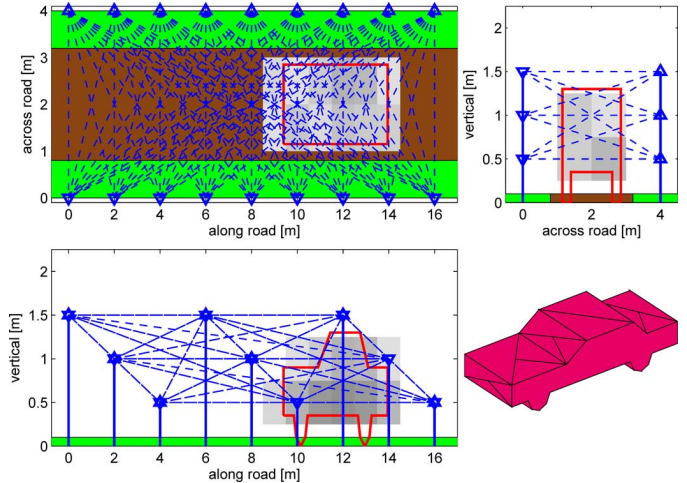

Fig. 1. Example of 3D RTI for roadside surveillance. The dashed lines indicate the network connectivity. The vehicle was a mustang and its true position is indicated by the solid outline. The estimated obstruction is indicated by the shading.

A calibration step is performed to determine the baseline RSS of each link. Later, the differences in RSS $\mathbf{y} \in \mathcal{R}^{M \times 1}$ due to differences in voxel densities $\mathbf{x} \in \mathcal{R}^{N \times 1}$ are [4]

$$\mathbf{y} = \mathbf{W}\mathbf{x} + \mathbf{n}, \qquad (1)$$

with weighting matrix $\mathbf{W} \in \mathcal{R}^{M \times N}$ and measurement noise $\mathbf{n} \in \mathcal{R}^{M \times 1}$. The drop in RSS in dBm on link $m$ is denoted $y_m$, the attenuation density of voxel $n$ is $x_n$, and a link's attenuation is a sum over voxel attenuation density times the weight value relating that link and voxel. In some instances when there is an obstruction but its density is not known numerically, $\mathbf{x}$ can be treated as a binary "occupancy" by comparing its real-valued entries to some threshold; however, we use the real-valued interpretation whenever possible.

The weighting matrix $\mathbf{W}$ has taken on various forms in the literature. In all cases, it can be decomposed as

$$\mathbf{W} = \mathbf{S} \odot \mathbf{\Omega} \qquad (2)$$

where $\mathbf{S}$ is a binary selection matrix, $\mathbf{\Omega}$ is a real-valued matrix containing the magnitudes of the weights, and $\odot$ indicates Hadamard (element-wise) multiplication.

Before discussing the various models for $\mathbf{S}$ and $\mathbf{\Omega}$, we need a few definitions. Let $d(m)$ be the length of link $m$; and let $d_1(m,n)$, $d_2(m,n)$ be the distances from the center of voxel $n$ to the two endpoints of link $m$. Let the tunable parameter $\lambda$ define an ellipse with the endpoints of link $m$ as the foci, as shown in Fig. 2(a); increasing $\lambda$ will increase the minor axis of the ellipse. Typical values of $\lambda$ are 0.1 to 0.01 feet. Let $L_{m,n}$ be the length of the segment of the link inside the voxel, as shown in Fig. 2(b).

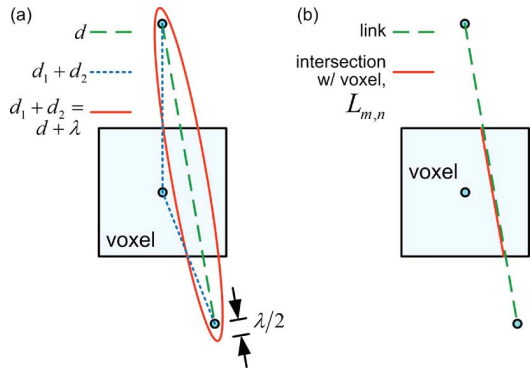

Fig. 2. Comparison of weight models. (a) Ellipse-based weight model. (b) Line integral based weight model.

Typical selection matrices include

$$S_{m,n}^{\text{Ellipse}} = \mathbb{1}\left[d_1(m,n) + d_2(m,n) < d(m) + \lambda\right] \quad (3)$$
$$S_{m,n}^{\text{Line}} = \mathbb{1}\left[\text{link } m \text{ traverses voxel } n\right] \quad (4)$$
$$S_{m,n}^{\text{All}} = 1 \quad (5)$$

where $\mathbb{1}[\cdot]$ is the indicator function. The Ellipse selection matrix was proposed by [4] and is the most common in the literature. The Line selection matrix was used in [12], [13], [17], [25], [26]; and all values of $\mathbf{W}$ were populated in [27] and [28], implicitly using a non-selective $\mathbf{S}$ matrix. The continuous portion of the weights, given by $\mathbf{\Omega}$, has been modeled as

$$\Omega_{m,n}^{\text{NeSh}} = (d(m))^{-1/2} \quad (6)$$
$$\Omega_{m,n}^{\text{Line}} = L_{m,n} \quad (7)$$
$$\Omega_{m,n}^{\text{NeShLine}} = L_{m,n} \cdot (d(m))^{-1/2} \quad (8)$$
$$\Omega_{m,n}^{\text{ExpDec}} = \exp\left(-\frac{\tilde{\lambda}_{m,n}}{(2\sigma_\lambda)}\right) \quad (9)$$
$$\Omega_{m,n}^{\text{InvArea}} = \left[\frac{\pi}{4}\left(d(m) + \tilde{\lambda}_{m,n}\right)\sqrt{2d(m)\tilde{\lambda}_{m,n} + \tilde{\lambda}_{m,n}^2}\right]^{-1} \quad (10)$$

where $\tilde{\lambda}_{m,n}$ is the value of $\lambda$ required to have the ellipse discussed above pass through the center of the voxel, and $\sigma_\lambda$ is a tunable parameter. The NeSh model was proposed by [4] and is again the most common in the literature. The Line model was used in [17], [26], and the hybrid NeSh-Line model was used in [12], [13]. In (9) and (10), Li *et al.* [27] and Hamilton [28] both modeled the weights as decreasing as the ellipse size increases, though [27] used an exponential decay model with tunable parameter $\sigma_\lambda$, whereas [28] modeled the weights as the inverse of the ellipse's area.

We favor the Line model for both $\mathbf{S}$ and $\mathbf{\Omega}$ since the attenuation should depend on the path length through the obstruction, and this model resembles the method used in computed tomography scans in medical imaging. It is also fairly simple to implement – while computing $L_{m,n}$ requires several lines of logic, it is computationally cheap. Approximating the voxels as ellipsoids for the sake of tractable theoretical analysis was discussed in [26].

The noise $\mathbf{n}$ is commonly modeled as Additive White Gaussian Noise (AWGN) [12], [13], [22], [27]. However, in [4], measured data was used to fit a Gaussian mixture model to the noise, with two zero-mean Gaussians with weights of 0.548 and 0.452 and standard deviations of 0.971 dB and 3.003 dB. We have found that an AWGN error model with a standard deviation $\sigma_n$ of 4 dB to 6 dB is a good fit for the experimental data obtained from USNA's and AFIT's RTI testbeds, and due to its analytic tractability, we will assume an AWGN model in this paper.

The obstruction $\mathbf{x}$ is typically modeled as a deterministic unknown. For example, [4] used a cylindrical human model,

$$x_n = \begin{cases} 1, & \|\boldsymbol{\chi}(n) - \mathbf{c}_o\| < r_o \\ 0, & \text{else} \end{cases} \quad (11)$$

where $\boldsymbol{\chi}(n)$ is the position vector of voxel $n$, $x_n$ is a binary quantity that in this case is more of an occupancy than a density, and $\mathbf{c}_o$ and $r_o$ are the center point and radius of the obstruction. For use in Bayesian estimation algorithms, it may be useful to model $\mathbf{x}$ as a random field. In [4] a Gaussian prior was used,

$$\mathbf{x} \sim \mathcal{N}(\mathbf{0}, \mathbf{C}_x)$$
$$[\mathbf{C}_x]_{n_1, n_2} = \sigma_x^2 e^{-d(n_1, n_2)/\delta_c} \quad (12)$$

where $d(n_1, n_2)$ is the distance between voxels $n_1$ and $n_2$, $\sigma_x^2 = 0.1 \text{ dB}^2$, and $\delta_c = 1.3$ m is the correlation "space constant." However, a Gaussian prior is problematic, since physically the obstruction is non-negative. Alternative, physical priors will be considered in Section IV-A.

To estimate $\mathbf{x}$, [4] used a least-squares solution with an additive Tikhonov regularization term,

$$\hat{\mathbf{x}} = \arg\min_{\mathbf{x}} \|\mathbf{W}\mathbf{x} - \mathbf{y}\|_2^2 + \alpha\,\mathbf{x}^T\mathbf{Q}\mathbf{x} \quad (13)$$
$$= \left(\mathbf{W}^T\mathbf{W} + \alpha\mathbf{Q}\right)^{-1}\mathbf{W}^T\mathbf{y}, \quad (14)$$
$$\mathbf{Q} \triangleq \mathbf{D}_x^T\mathbf{D}_x + \mathbf{D}_y^T\mathbf{D}_y + \mathbf{D}_z^T\mathbf{D}_z, \quad (15)$$

where $\mathbf{D}_d$ computes the derivative in dimension d (see [18] for a detailed example) and $\alpha$ is a user-determined constant. Note that if the noise is Gaussian, $\alpha = 0$ yields the Maximum Likelihood (ML) estimate of $\mathbf{x}$. However, without the regularization term, the matrix $\mathbf{W}^T\mathbf{W}$ is typically not full rank. When $\alpha > 0$, it is experimentally-determined, and it indicates the relative emphasis on the regularization term.

## III. HARDWARE DESIGN

We now describe the wireless sensor network used in this work and the layout of its nodes. Additionally, we provide a description of networked RTI operation.

### A. Sensor Platform

Our RTI sensor network is comprised of a number of custom-designed wireless sensor nodes and one Command and Control (C2C) node. A block diagram of the sensor node is given in Fig. 3; the hardware for all nodes is identical, although the C2C node uses a different software code. Each node consists of a battery power source, PIC microcontroller, and XBee ZigBee Pro RF module. ZigBee devices are widely used in a variety of WSNs, and were selected as they provide a low-cost/low-power platform along with a simplified protocol stack. For operation in the RTI network, the XBee is configured to transmit on one of the sixteen 2 MHz wide ZigBee channels in the 2.4 GHz ISM band. The XBee is also configured to use Advanced Peripheral Interface (API) mode to ensure consistent packet formats. An RS-232 port provides an off-board serial interface for communication with a computer or other device. An example of a sensor node is shown in Fig. 4, which includes both the node and a metal backplane to reduce the effects of multipath on the measured RSS value (discussed further in Section III-D).

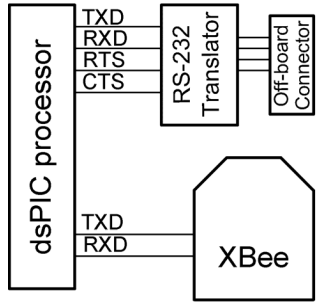

Fig. 3. Block diagram of the USNA RTI sensor node.

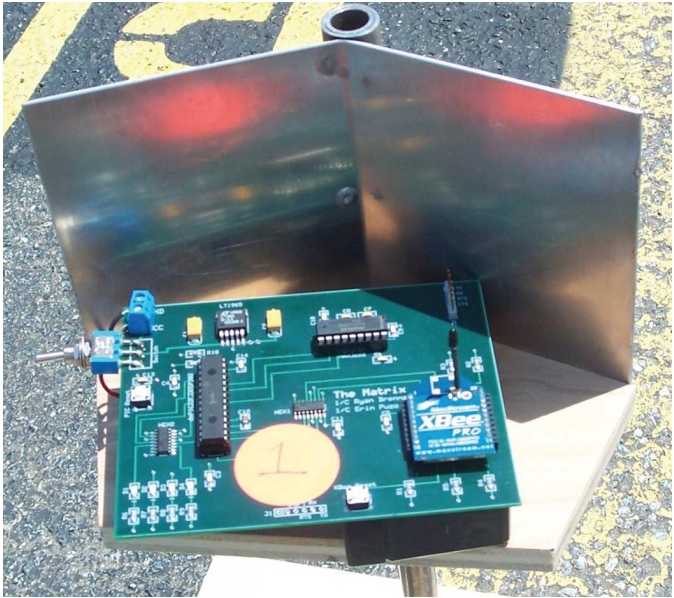

Fig. 4. RTI node from USNA testbed.

Metal backplanes, however, can distort the antenna pattern, potentially introducing destructive interference and deep nulls in the pattern. Therefore, the pattern of the entire sensor node, including backplane, was measured in the USNA Antenna Chamber. The Chamber uses a Diamond Engineering 6100 antenna positioning system coupled with an Anritsu MS4642A VectorStar Vector Network Analyzer. To perform the measurement, an XBee node was deconstructed (i.e., all RF hardware was removed, leaving only the antenna), and a coaxial cable was carefully soldered onto an RF input port. The antenna pattern was recorded at 2.45 GHz, and is shown in Fig. 5. From the figure, we observe no deep nulls across the main lobe of the pattern, and that the backplane significantly attenuates signals that would arrive from outside the RTI network.

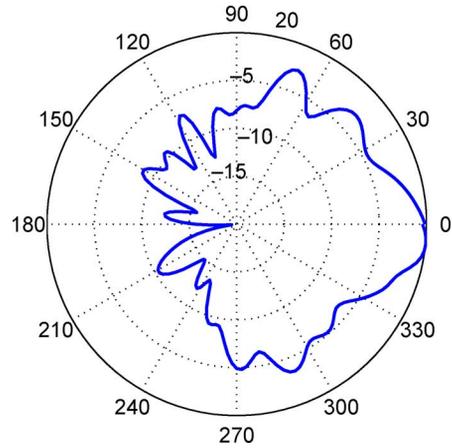

Fig. 5. Gain pattern of the node in Fig. 4, including the metal backplane. The backplane attenuates multipath from behind the sensor, focusing the energy on the lines of sight to the sensors across the road.

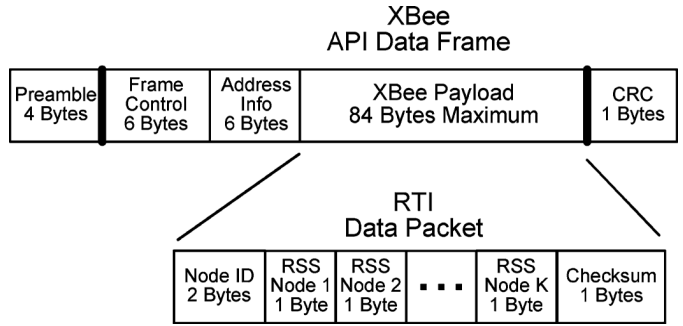

Fig. 6. The packet structure for the USNA RTI network as a subset of the ZigBee standard Data Frame.

### B. WSN Operation

A RTI scan of the network is initiated by a start packet transmitted by the C2C node. The start packet serves as a coarse synchronization of the overall network timing. Each node is assigned a specific time slot when it is allowed to transmit, with each slot defined as a time delay relative to the time when the node receives the start packet. Due to variation in propagation and processing delays across the network, nodes will receive the start packet at slightly different times. Rather than perform a fine synchronization or timing alignment step, sufficient guard time is built into each time slot to account for the timing variability across the network, albeit at the cost of a slower network scan time.

During each scan, nodes transmit sequentially in a token passing scheme; during its allocated time slot, each node will broadcast a single XBee API Data Frame Packet (Fig. 6). The remaining nodes in the network are configured to receive the packet and measure the RSS value. In the RTI Data section of the packet, each node transmits its unique 16-bit node identity (Node ID), followed by the most recent RSS values it has recorded from the prior scan (one RSS value for every other node in the network). Finally, for data verification purposes, the RTI section contains a simple checksum that validates the RSS data, while the Data Frame Packet as a whole is validated by a simple CRC check.

An entire network scan consists of each individual node transmitting its data once. At the completion of the scan, each node will have knowledge of the most recent signal strength between itself and every other node in the network array. During the scan, the C2C node will listen to all node transmissions, allowing it to record the RSS information in real time.

*C. Object Resolution*

For object detection and classification using RTI, an important consideration is the ability to unambiguously detect an object passing through the network. Unambiguous detection here means that the object should not cross a voxel boundary during a single scan of the network; i.e., the object does not move a distance greater than the length of one voxel during one scan. An object that is moving faster than this maximum allowed speed will appear "blurred" in the RTI data, which is analogous to photographing a fast-moving object with a shutter time that is too slow. As a result, the maximum unambiguous scan time supported by the network establishes an upper bound for the maximum speed that a vehicle can be traveling through the monitored area.

The unambiguous scan time is given by

$$T_{\text{scan}} = \left(\frac{L_{\text{RTI}}}{r_z} + T_g + T_p\right) K \quad (16)$$

where $T_{\text{scan}}$ is the amount of time in seconds required for a full scan of the network to occur, $L_{\text{RTI}}$ is the total number of bits in each Data Frame Packet, $K$ is the number of sensor nodes in the network, $r_z$ is the *baseband* rate[1] at which the data is transmitted from processor to the XBee in bits/second. Additionally, $T_g$ is the guard time for each slot in the token passing scheme and $T_p$ is the amount of time required by a node to process each packet.

In a linear configuration, where nodes line both sides of a roadway, the minimum desired object resolution will be inversely proportional to the maximum speed of the vehicle, as given by

$$v_{\text{det}} = \frac{r_{\text{dist}}}{T_{\text{scan}}} \quad (17)$$

where $r_{\text{dist}}$ is the voxel size and $v_{\text{det}}$ is the maximum movement velocity in meters/second at which objects moving through the network can be detected.

Similarly, the maximum velocity that can be *detected* is given by the total road length within the network divided by the scan time. Using 9 nodes on each side of the road with a 3 m spacing, the road length is $(9-1) \cdot 3 = 24$ m. Thus, a 1 s scan time, which is a simple goal for a field-ready system, leads to a maximum detectable velocity of 24 m/s (86 km/hour or 54 miles/hour).

*D. Practical Implementation Issues*

Practical RTI networks have three major issues to contend with: (i) multipath signals that may distort the true RSS values (ii) the relationship between scan time, voxel size, and monitored area, and (iii) the maximum number of nodes that can be installed in the network.

[1]Although ZigBee transmits at an RF data rate of $250 kbps$, the XBee modules interface with the processor at standard RS-232 data rates. Thus, even though packets are transmitted at the RF data rate, the system is ultimately limited by the baseband rate of the processor-to-XBee interface.

A crucial consideration for accurate object identification and classification is multipath mitigation. In cluttered roadside environments, multipath signals that reflect from the road surface or large objects in the nearby vicinity (such as buildings, trees, etc.) will perturb the RSS measurement by a node. These perturbations may result in an RSS value that is either higher or lower than would be expected if a direct-path only signal was present. One way of reducing or eliminating multipath signals is to construct a backplane behind each node, as shown in Fig. 4. The backplane prevents large reflective objects from outside of the network from influencing the RSS value measured by a node. Furthermore, because the backplane is reflective, precise positioning of the ZigBee transmitter antenna at the focal point will result in additional forward gain, similar to a corner reflector antenna.

The second consideration for accurate object identification is the tradeoff between resolution and monitored area. Roughly speaking, an RTI network with $K$ nodes will have $\mathcal{O}(K^2)$ unique links, resulting in $\mathcal{O}(K^2)$ voxels. This is because even through the estimation is regularized, the number of unknowns (voxels) should not be much more than the number of measurements (links). For a roadside scenario, where nodes only line the two sides of the road, the number of unique links, and consequently the number of voxels, decreases to $\mathcal{O}\left((K/2)^2\right)$. If all nodes are positioned at a single height, the RTI image will have voxels that span the width and height of the network, but each of which has a length equal to

$$\delta_x = \frac{2 D_{Node}}{K}, \quad (18)$$

where $D_{Node}$ is the spacing between RTI nodes on the same side of the road. If nodes are placed at $H_{Node}$ different heights, the RTI image will now have $H_{Node}^2$ voxels that span the width and height of the network, each of which has a length of

$$\delta_x = \frac{D_{Node} \times H_{Node}^2}{2K}. \quad (19)$$

Note that, for a given monitored area size, placing nodes at different heights results in a higher resolution RTI image, but at the cost of an increased density of nodes.

The final consideration is the maximum number of nodes that can be supported by the RTI WSN. For a network that transmits a single packet per node per scan, the maximum number of nodes in the network is limited by the size of the packet. The ZigBee specification allows a maximum Data Frame packet size of 104 bytes [29]; the XBee devices used in our network have a maximum data payload of 84 bytes in normal mode, or 255 bytes in API fragmentation mode [30]. For our RTI packet structure, 3 bytes are reserved for overhead (Node ID and a data checksum) and 1 byte is required per RSS measurement. Thus, assuming one packet transmitted per scan, our network could support a maximum of 81 nodes in normal mode or 252 nodes in packet fragmentation mode. The XBee devices have a practical baseband data rate upper limit of approximately 38.4 kbps [30], which (assuming no guard time) results in an 81-node network best-case scan time of 1.75 seconds.

To expand the maximum number of nodes in the network or improve the scan time, the simplest solution is to break the overall network up into a set of sub-networks. Each sub-network would then operate on a different ZigBee RF channel. The sub-networks could either be interleaved within the coverage area (to increase voxel density in the coverage area) or physically separated using frequency reuse (to increase the coverage area for a given voxel size). In either case, an additional level of network hierarchy would need to be created, where the C2C nodes would report their subset of data upstream to a data aggregator node. The data aggregator(s) would combine RSS data from all sub-networks into a complete RTI data scan. Furthermore, the sub-network concept could be used to improve the network scan time for a smaller sized network; with a small subset of nodes on each of the sub-networks, the the scan time would be improved by a factor equal to the number of sub-networks.

An additional option to expand the number of nodes in the network is to simply transmit multiple packets per network scan, at the cost of scan time and some additional packet overhead in the form of a sequence ID field. The multi-packet approach is convenient from an implementation standpoint, as it simply requires a software change to the network—no additional physical planning or network devices are required. As an example, for the USNA network, the 2-byte Node ID field would allow for a maximum of 65,535 nodes, requiring 840 packets per network scan and a 2-byte Sequence ID field.

## IV. Estimation Algorithms

It is known that regularization is equivalent to assuming a Gaussian Bayesian prior on $\mathbf{x}$. In Section IV-A we show how this leads to a natural choice of $\alpha$, and we extend this idea by deriving Maximum A Posteriori (MAP) estimators for more realistic choices of the prior $f(\mathbf{x})$. In Section IV-B we also investigate methods for dealing with aphysical situations in which some entries in $\mathbf{y}$ are negative, which can provide negative estimates for the obstruction $\hat{\mathbf{x}}$. Lastly, Section IV-C discusses estimation algorithms for moving vehicles, which enables the use of multiple frames of data to form a single scene estimate.

### A. Incorporating Physical Priors

First, recall the AWGN fading model, and consider a Gaussian Bayesian prior on $\mathbf{x}$, so that

$$f_{\text{awgn}}(\mathbf{y}|\mathbf{x}) \sim \mathcal{N}\left(\mathbf{W}\mathbf{x}, \sigma_n^2 \mathbf{I}_M\right) \quad (20)$$
$$f_{\text{gaussian}}(\mathbf{x}) \sim \mathcal{N}(m\mathbf{1}_{N\times 1}, \mathbf{C}_x), \quad (21)$$

where $\mathbf{I}_M$ is the $M \times M$ identity matrix and $\mathbf{C}_x$ was defined in (12). The mean $m$ is typically zero, but is left generic here. The corresponding MAP estimator is

$$\begin{aligned}
\hat{\mathbf{x}}_{\text{MAP}} &= \arg\max_{\mathbf{x}} f(\mathbf{x}|\mathbf{y}) & (22) \\
&= \arg\max_{\mathbf{x}} f(\mathbf{y}|\mathbf{x}) f(\mathbf{x}) \\
&= \arg\max_{\mathbf{x}} \left( \frac{-\|\mathbf{W}\mathbf{x}-\mathbf{y}\|^2}{2\sigma_n^2} - \frac{1}{2}(\mathbf{x}-m\mathbf{1})^T \mathbf{C}_x^{-1}(\mathbf{x}-m\mathbf{1}) \right) \\
&= \arg\min_{\mathbf{x}} \left( \|\mathbf{W}\mathbf{x}-\mathbf{y}\|^2 + (\mathbf{x}-m\mathbf{1})^T (\sigma_n^2 \mathbf{C}_x^{-1})(\mathbf{x}-m\mathbf{1}) \right) \\
&= \left(\mathbf{W}^T\mathbf{W} + \sigma_n^2 \mathbf{C}_x^{-1}\right)^{-1}\left(\mathbf{W}^T\mathbf{y} + m\sigma_n^2 \mathbf{C}_x^{-1}\mathbf{1}\right). & (23)
\end{aligned}$$

If $m = 0$ as in (12) and if $\alpha \mathbf{Q} = \sigma_n^2 \mathbf{C}_x^{-1}$, then this is equivalent to the regularized ML solution (14). This fact will be used to derive an appropriate value for $\alpha$ without trial-and-error. However, if $m > 0$, then the algorithm of (14) is no longer optimal in the MAP sense, and the modified algorithm shown in (23) should be used.

Assuming $m = 0$ and considering just one dimension for simplicity,

$$\sigma_n^2 \mathbf{C}_x^{-1} = \frac{\frac{\sigma_n^2}{\sigma_x^2}}{1-c^2} \begin{bmatrix} 1 & -c & 0 & 0 & \cdots & 0 \\ -c & 1+c^2 & -c & 0 & \cdots & 0 \\ 0 & -c & 1+c^2 & -c & \cdots & 0 \\ \vdots & \vdots & \cdots & \ddots & \ddots & \vdots \\ 0 & 0 & 0 & 0 & \cdots & 1 \end{bmatrix} \quad (24)$$

where $c = e^{-\delta/\delta_c} = 0.9260$ for the parameters in [4]. To compare, in one dimension, if we regularize by averaging the effects of forwards and backwards differences,

$$\mathbf{Q} = \frac{1}{2}\left(\mathbf{D}_{x,fwd}^T \mathbf{D}_{x,fwd} + \mathbf{D}_{x,bck}^T \mathbf{D}_{x,bck}\right) \quad (25)$$

$$= \begin{bmatrix} 1.5 & -1 & 0 & 0 & \cdots & 0 \\ -1 & 2 & -1 & 0 & \cdots & 0 \\ 0 & -1 & 2 & -1 & \cdots & 0 \\ \vdots & \vdots & \cdots & \ddots & \ddots & \vdots \\ 0 & 0 & 0 & 0 & \cdots & 1.5 \end{bmatrix} \quad (26)$$

Thus, $\alpha \mathbf{Q} \approx \sigma_n^2 \mathbf{C}_x^{-1}$ if

$$\boxed{\alpha = \frac{\sigma_n^2}{\sigma_x^2} \cdot \frac{1}{1-c^2} \approx \frac{\sigma_n^2}{\sigma_x^2}\frac{\delta_c}{2\delta}} \quad (27)$$

This choice of $\alpha$ makes (14) equivalent to the MAP estimator, provided (12) is a reasonable choice of prior.

Note that $\sigma_x^2/\sigma_n^2$ is a Signal to Noise Ratio (SNR) and $\delta_c/\delta$ is the number of voxels of separation required for 63% decorrelation. Using the parameter values from [4] in (27) suggests $\alpha \approx 0.3$ for weak fading up to $\alpha \approx 30$ for strong fading. The experimental optimum in [4] was $\alpha \approx 5$, with good values in the range [2, 20]. Thus, the MAP approach provides a good method for selecting $\alpha$.

It is interesting to consider the effective correlation matrix $\mathbf{C}_x$ that would make $\alpha \mathbf{Q} = \sigma_n^2 \mathbf{C}_x^{-1}$ exactly true. Inverting (26) yields the $\mathbf{C}_x \propto \mathbf{Q}^{-1}$ shown in Fig. 7, which is no longer Toeplitz as in (12). Thus, if Tikhonov regularization is used, voxels near the middle of the WSN are effectively given an a priori bias to have more energy than those nearer the edges.

Hitherto, we have considered the Gaussian prior on $\mathbf{x}$, usually with a zero mean. Recall that $\mathbf{x}$ is the drop in RSS after an obstruction enters the scene; thus, barring unusual multipath effects, $\mathbf{x}$ should be positive for voxels with a new obstruction, zero for voxels for no change, and negative only in voxels where an obstruction was present during calibration but was subsequently removed. As such, a Gaussian prior is a poor choice. The frequency of negative values can be reduced by increasing $m$, but that will also make small values of $\mathbf{x}$ less frequent.

A realistic yet tractable prior distribution for $\mathbf{x}$ could be

$$f_{\exp}(\mathbf{x}) = \Pi_{n=1}^N m^{-1}\exp\left(-\frac{x_n}{m}\right), \quad x_n \geq 0\,\forall n. \quad (28)$$

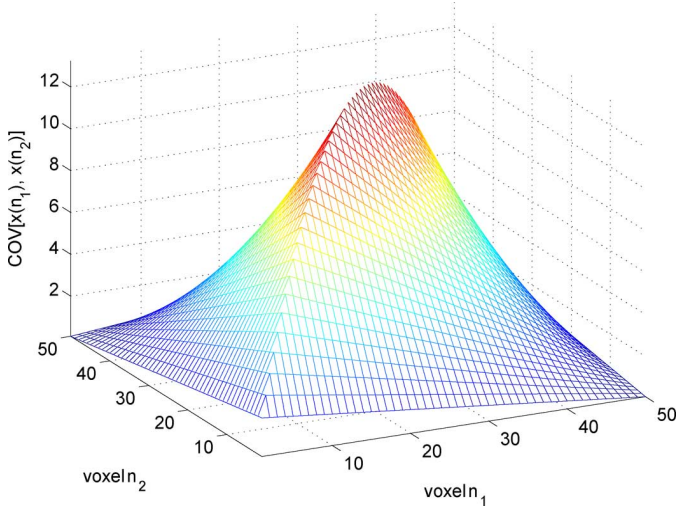

Fig. 7. Effective covariance matrix of Bayesian prior due to Tikhonov regularization. For simplicity, only a 1D scene is depicted here.

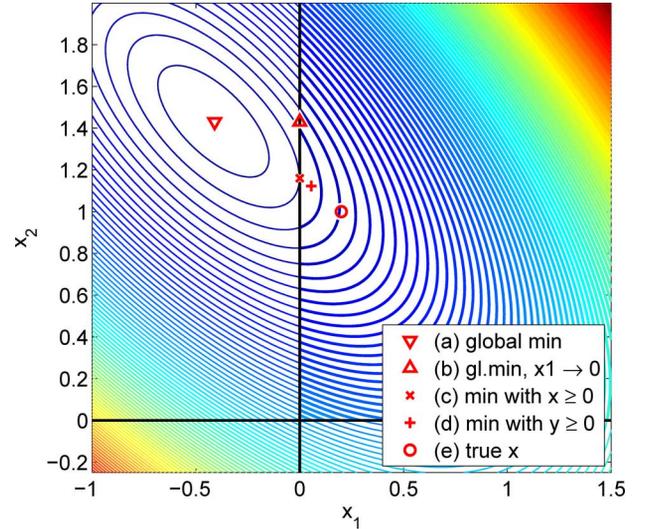

Fig. 8. Solutions for dealing with negative values of RSS drop observations $y$, which lead to negative values of voxel estimates $\hat{x}$.

However, the exponential distribution does not generalize well to the multivariate case unless independence is assumed, which may not be realistic. The exponential prior with an independence assumption leads to the MAP estimator

$$\hat{\mathbf{x}}_{\text{MAP}} = \arg\max_{\mathbf{x}} f(\mathbf{y}\,|\,\mathbf{x})f(\mathbf{x}) \quad (29)$$

$$= \arg\max_{\mathbf{x},x_n \geq 0} \left( \frac{-1}{2\sigma_n^2} \|\mathbf{W}\mathbf{x} - \mathbf{y}\|^2 - \frac{1}{m}\mathbf{x}^T\mathbf{1} \right)$$

$$= \arg\min_{\mathbf{x},x_n \geq 0} \left( \|\mathbf{W}\mathbf{x} - \mathbf{y}\|^2 + \left(\frac{2\sigma_n^2}{m}\right)\mathbf{x}^T\mathbf{1} \right). \quad (30)$$

Provided it lies in the realm of non-negative values of $\mathbf{x}$, the corresponding solution is

$$\hat{\mathbf{x}}_{\text{MAP}} = \left(\mathbf{W}^T\mathbf{W}\right)^{-1}\left(\mathbf{W}^T\mathbf{y} - \left(\frac{\sigma_n^2}{m}\right)\mathbf{1}\right). \quad (31)$$

However, that is rarely the case due to noise. The next section will address methods of constraining $\mathbf{x}$ and/or $\mathbf{y}$ to the first (all-positive) quadrant of the parameter space. Note that (30) is almost a "basis pursuit" problem [31], which minimizes $\|\mathbf{x}\|_1$. Here, however, there are no absolute values on the elements of $\mathbf{x}$, and instead they are constrained to be non-negative.

In summary, we have three candidate MAP solution algorithms. Using the zero-mean Gaussian prior as in (12) leads to the Tikhonov-regularized least squares solution, with $\alpha$ chosen according to (27). Using a Gaussian prior with mean $m > 0$ leads to the solution (23); this reduces the preference of the solution of negative $\mathbf{x}$ at the expense of also reducing its preference for near-zero $\mathbf{x}$. Finally, the i.i.d. exponential prior leads to the solution of (31). Making use of the fact that $\mathbf{C}_x^{-1}\mathbf{1} \propto \mathbf{1}$, all of these solutions together are of the generic form

$$\hat{\mathbf{x}}_{\text{MAP}} = \left(\mathbf{W}^T\mathbf{W} + \alpha\mathbf{Q}\right)^{-1}\left(\mathbf{W}^T\mathbf{y} + \beta\mathbf{1}\right), \quad (32)$$

where $\alpha \geq 0$ but $\beta$ can be positive or negative, and it is assumed that all $\hat{x}_n$ are non-negative. Illustration of the relative effects of $\alpha$ and $\beta$, along with methods for dealing with negative values of $y_m$ and/or $x_n$, are the discussed in the next subsection.

### B. Dealing With Negative Observations

Often, negative entries are observed in $\mathbf{y}$. This can occur due to inadequate calibration of the empty network or due to large amounts of noise. Fig. 8 shows the search space for a toy problem of 2 voxels side by side, with 2 nodes above and 2 nodes below, leading to 4 links total. If some link has a large enough negative value, the contours of the cost surface of (13) are as in Fig. 8, with the global minimum in the undesirable realm of $x_1 < 0$. If we simply compute (a), the solution of (14), and replace the negative value of $\hat{x}_1$ by zero, we get a poor solution, (b). However, if we modify the Gaussian prior on $\mathbf{x}$ to be a truncated Gaussian allowing only positive $x$ values, or simply use the already-truncated exponential prior, the contours should be minimized over the space $x_n \geq 0 \forall n$, i.e., the first quadrant in Fig. 8. The minimum within the constrained space is a much better solution, (c). Alternatively, we could force $y$ to be positive by replacing all negative entries by zero, leading to a comparable solution, (d); however, it is more justifiable to constrain the search space than to modify the observed data, and the exponential prior requires a constrained search regardless. The true $\mathbf{x}$ in this example is (e). Although (c) and (d) lead to similar solutions in this example, for more complex problems, (c) results in a better solution, as will be shown later in this section.

Solving the generic problem $\min_{\mathbf{x},x_n \geq 0} \|\mathbf{W}\mathbf{x} - \mathbf{y}\|^2$ is quite common, and solutions go by the name Non-Negative Least Squares (NNLS), and the algorithm in [32, Chapter 23, p. 161] is widely used. However, NNLS generally do not incorporate a regularization term; most algorithms either allow for regularization or a constraint, but rarely both. A notable exception is the Projected Gradient Method (PGM) [33], which solves problems of the form

$$\min_{\mathbf{x} \in \mathcal{S}} f(\mathbf{x}) \quad (33)$$

using iterations of

$$\hat{\mathbf{x}}_{k+1} = P\left[\hat{\mathbf{x}}_k - \mu \nabla_{\mathbf{x}} f(\mathbf{x}_k)\right] \quad (34)$$

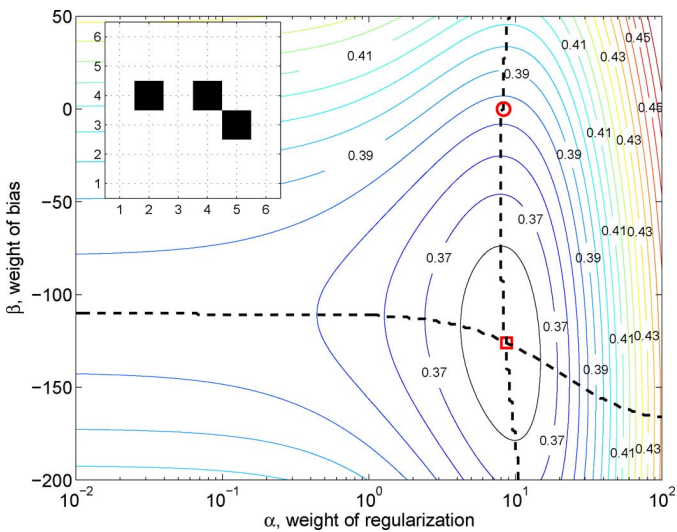

Fig. 9. RMSE$(\alpha, \beta)$ when negative entries of **y** are set to zero before solving. Legend: ○ is the commonly used solution from [4] (RMSE = 0.385), □ is the solution optimized over $(\alpha, \beta)$ (RMSE = 0.365), and the dashed lines are the solutions optimized by fixing $\alpha$ or $\beta$ and optimizing the other parameter. The inset shows the simulated scene, which used 28 sensors spaced around the perimeter.

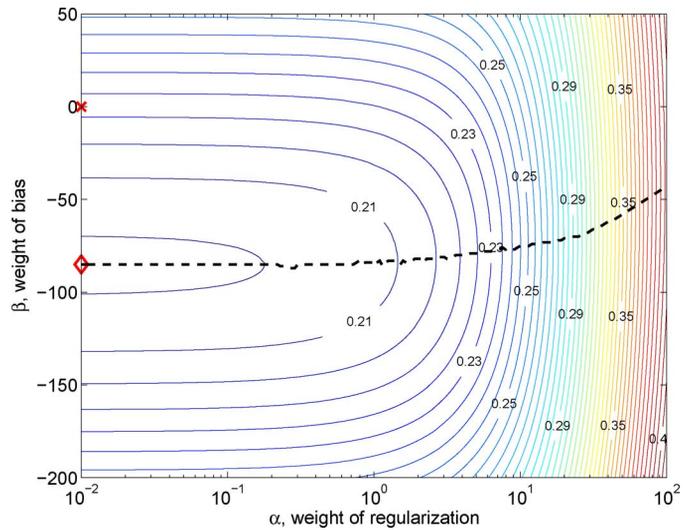

Fig. 10. RMSE$(\alpha, \beta)$ when the solution is constrained to have all $\mathbf{x} \geq 0$. Legend: × is the solution analogous to [4] with the $\mathbf{x} \geq 0$ constraint added (RMSE = 0.221), ◇ is the solution optimized over $(\alpha, \beta)$ (RMSE = 0.204), and the dashed line shows the solutions optimized by fixing $\alpha$ and optimizing over $\beta$.

where $P[\cdot]$ projects its argument onto the set $\mathcal{S}$. In our case, $f(\mathbf{x})$ is given by the argument of (13), and the constraint set is $\mathcal{R}_+^N$, the set of non-negative **x**.

The main drawback of PGM is that it is computationally expensive; in our problem, computing the gradient requires $N^2$ multiplications per iteration, with many iterations (on the order of 50 for our results in Section V) needed for convergence. We propose a similar approach that is computationally attractive when **x** is expected to be sparse. Note that a tentative solution can be found by solving once to determine the problem elements of **x** (i.e., those that are estimated to be negative), setting them to zero, then resolving only for the remaining elements of **x**. That is, we replace the gradient step with an exact solution, and after each projection we fix the elements affected by the projection to zero for the rest of the iterations (removing them from any subsequent computational burden).

Thus, the proposed method to approximate solution (c) is:
1) Compute $\hat{\mathbf{x}}$ via (14).
2) Find the negative elements of $\hat{\mathbf{x}}$ and omit the corresponding columns of **W** and of each $\mathbf{D}_d$.
3) Repeat (14) using the reduced matrices.
4) Iterate through steps 2–3 if necessary.

The iteration is needed because with more than 2 voxels, the cost surface can be more complex than depicted in Fig. 8. The first iteration usually removes most of the negativity, but a few elements in the re-solved (14) may become negative in the next iteration. For our results in Section V, 3 iterations (counting the initial estimate in step 1) appear to be sufficient.

To illustrate the two classes of methods of dealing with negative data (setting negative $y_m$ to zero or constraining $x_n \geq 0$) in conjunction with the general solution of (32), consider the 36-voxel scene shown in the inset of Fig. 9, with 28 sensors around the perimeter, similar to the test scene in [4]. Fig. 9 shows the Root Mean Squared Error (RMSE) contours as functions of the regularization weight $\alpha$ and the bias weight $\beta$, when negative $y_m$ are set to zero before solving with (32). Similarly, Fig. 10 shows contours when negative $y_m$ are allowed but negative $x_n$ are mitigated via the 4-step algorithm proposed early in this subsection. The combination of optimizing the bias weight $\beta$ and constraining $x_n \geq 0$ reduces the RMSE from 0.385 to 0.204 relative to the traditional approach in [4], nearly halving the error. The majority of the improvement comes from constraining the search space, though the bias term helps incrementally.

Also note that the process of removing the negative values of $x_n$ from the search space has made the problem better conditioned, greatly reducing the need for the $\alpha \mathbf{Q}$ regularization, as the contours in Fig. 10 show that for any given $\beta$, letting $\alpha \to 0$ monotonically improves the solution; though the term is needed for the first iteration before any voxels are removed from the search.

### C. Moving Vehicles

The scenario that we are primarily interested in is roadside surveillance. Specifically, we are interested in placing sensors along a roadway to image vehicles moving along the roadway, as shown in Fig. 1. Assuming that the vehicle is moving at a constant (but unknown) velocity while in the network, it is possible to combine multiple frames of data to produce a single vehicle image estimate, and simultaneously estimate the velocity. This is somewhat analogous to the problem of resolving moving targets in synthetic aperture radar imagery, though since RTI operates on power data only, we have no issues with phase coherence.

Define the dimensions $x, y, z$ to be along the road, across the road, and vertical, with $N_x$, $N_y$, and $N_z$ voxels in those respective dimensions; cf. Fig. 1. Let $\mathbf{J}_0$ be a square $N_x \times N_x$ matrix with ones on the first subdiagonal and zeros elsewhere. Though it is not strictly invertible, for notational convenience define $\mathbf{J}_0^{-1}$ to be the transpose of $\mathbf{J}_0$, i.e., with ones on the first superdiagonal. Left multiplication of a column vector by $\mathbf{J}_0^d$ or $\mathbf{J}_0^{-d}$ has the effect of shifting it by $d$ elements down or up, respectively, and padding the end with $d$ zeros. Now let $\mathbf{J} = \mathbf{I}_{N_y N_z \times N_y N_z} \otimes \mathbf{J}_0$, where $\otimes$ is the Kronecker product. The expanded shifting matrix $\mathbf{J}$ acts to shift a scene $\mathbf{x}$ left (negatively) along the road; similarly $\mathbf{J}^{-1}$ shifts the scene right (positively) along the road.

TABLE I
EXPERIMENTAL PARAMETERS. THE VEHICLES WERE A MUSTANG (M),
DELIVERY VAN (V), SMALL ELECTRIC CAR (E), AND SCHOOL BUS (B)

| Test | 1 | 2 | 3 | 4 | 5 | 6 | 7 | 8 | 9 |
|---|---|---|---|---|---|---|---|---|---|
| vehicle | m | v | e | b | m | m | m | m | m |
| # frames | 5 | 5 | 4 | 4 | 5 | 12 | 12 | 21 | 21 |
| $v$ [m/frame] | 4 | 4 | 4 | 4 | 4 | 1 | 1 | 1 | 1 |
| pole sep. [m] | 2 | 2 | 2 | 2 | 1 | 1 | 1 | 2 | 2 |
| reflectors? | y | y | y | y | y | n | y | n | y |

If the velocity $v$ is such that the vehicle moves an integer number of voxels per observation frame[2], then the concatenation of multiple frames can be written as

$$\underbrace{\begin{bmatrix} \mathbf{y}^1 \\ \mathbf{y}^2 \\ \mathbf{y}^3 \end{bmatrix}}_{\mathbf{y}_{\text{stack}}} = \begin{bmatrix} \mathbf{W}\mathbf{J}^{-v}\mathbf{x} \\ \mathbf{W}\mathbf{x} \\ \mathbf{W}\mathbf{J}^{v}\mathbf{x} \end{bmatrix} + \underbrace{\begin{bmatrix} \mathbf{n}^1 \\ \mathbf{n}^2 \\ \mathbf{n}^3 \end{bmatrix}}_{\mathbf{n}_{\text{stack}}} \quad (35)$$

where $v$ is expressed in units of voxels per frame. If necessary, $v$ and its multiples are rounded to integers. Here we are using three frames for simplicity, and we always shift relative to the center frame since the entire vehicle is not usually visible in the first frame. This approach allows us to solve for a single image $\mathbf{x}$ rather than one per frame,

$$\mathbf{y}_{\text{stack}} = \mathbf{W}_{\text{stack}}\mathbf{x} + \mathbf{n}_{\text{stack}}, \quad (36)$$

$$\mathbf{W}_{\text{stack}} = \begin{bmatrix} \mathbf{W}\mathbf{J}^{-v} \\ \mathbf{W} \\ \mathbf{W}\mathbf{J}^{v} \end{bmatrix} \quad (37)$$

which can be solved by any of the methods discussed in this paper, such as (32). Since $v$ is unknown, (36) must be solved for a range of tentative values of $v$. The best velocity estimate can be chosen via

$$\widehat{v}_{\text{ML}} = \arg\min_v \|\mathbf{W}_{\text{stack}}\widehat{\mathbf{x}}(v) - \mathbf{y}_{\text{stack}}\|^2. \quad (38)$$

## V. FIELD TESTS

To evaluate the operation of the roadside RTI WSN, a series of nine unique tests were conducted in a large open parking lot using a set of $K = 18$ WSN nodes and one C2C node in a configuration similar to that illustrated in Fig. 1. The experimental parameters are listed in Table I. Nodes were set up in two parallel rows with a distance of 4 meters in between rows. Nodes on a particular row were spaced a distance of 1 or 2 meters apart (see Table I), and the C2C node was stationed at one end of the network. Sensor nodes were installed on poles in a repeating "sawtooth" pattern at heights of $\{0.5, 1.0, 1.5\}$ meters. The sawtooth pattern was chosen as it provides a reasonable coverage area along with a relatively high density of voxels.

[2]If this assumption is undesirable, $\mathbf{J}$ can be augmented to be a fully-populated sub-voxel shifting matrix at the cost of additional computational complexity.

Testing was performed with four different vehicles: a small two-person electric car, a large-size passenger car (mustang), a large cargo van, and a large school bus. However, in most figures below we focus on the mustang, since the other vehicles tended to block either a tiny or a large portion of the network, leading to less distinctions between algorithms than the more challenging mid-sized vehicle. Prior to each test, a set of 20 calibration scans were recorded with no obstructions inside the monitored area. In order to have "truth" data, the tests were semi-static, with the vehicle moved to a precise position for each frame. To simulate motion as accurately as possible, the number of frames was as high as 21, with as little as 1 m movement between frames. We have also performed 2 tests imaging rolling vehicles with similar image quality, but due to the difficulty with comparing to truth data, those results are not presented here.

For our configuration, we were able to achieve 102 total voxels in a given RTI image, in line with the nominal $\mathcal{O}\left((K/2)^2\right) = 81$ voxels we would expect from using $K = 18$ sensor nodes. Again, this is because even with regularization, the number of unknowns should not significantly exceed the number of measurements. Using three node heights, we were able to achieve a vertical resolution of $N_z = 3$ voxels. Additionally, using the motion constraint we combined data from all RTI scans to increase the voxel count to 396, for an improvement of 4.9 over the nominal case.

Results will be discussed in the next few subsections. Note that the entire 3D scene was estimated, but only 2D side views are shown here for display purposes. The results are separated into (i) qualitative image quality, (ii) Automatic Target Recognition (ATR) performance, (iii) Receiver Operating Characteristic (ROC) curve comparison of the algorithms and of using or omitting the metal reflectors. These goals reflect the fact that we are specifically interested in either improving a human observer's ability to perform target recognition or in improving a computer's ability to automatically perform the same task. The effects on human-performed target recognition will be assessed indirectly both qualitatively, via imagery, and quantitatively, via ROC curves from segregating the voxels into areas that are and are not part of a vehicle. Ultimately, though, the goal of this section is to improve some form of target recognition rather than to track motion, since the problem of motion tracking of a vehicle on a straight road is rather trivial.

### A. Environmental Issues

Although the data presented in this section was recorded under favorable weather and environmental conditions, we note that weather and atmospheric effects have only a small impact on the received signal strength at 2.4 GHz.

Many weather and atmospheric propagation models, such as the ITU-R models for Rain, Cloud Absorption, and Gaseous Absorption, have very low specific attenuations at 2.4 GHz, typically much less than 1 dB/km [34]–[36]. For example, at a rain rate of 100 mm/hr, the specific attenuation at 2.4 GHz is only 0.03 dB/km, the attenuation due to clouds (e.g., fog) is less than 0.05 dB/km, and the attenuation due to atmospheric gasses is approximately 0.02 dB/km. For our configuration, our maximum link distance was approximately 38 meters. Operating in the conditions described above, the excess attenuation due to weather would be less than 0.005 dB. For a configuration which monitored a 100 m × 10 m road, the maximum link distance would be 110 m and the excess attenuation due to weather would be 0.011 dB, which is well within the ±4 dB error bounds in RSS measurement allowed by the ZigBee specification [29].

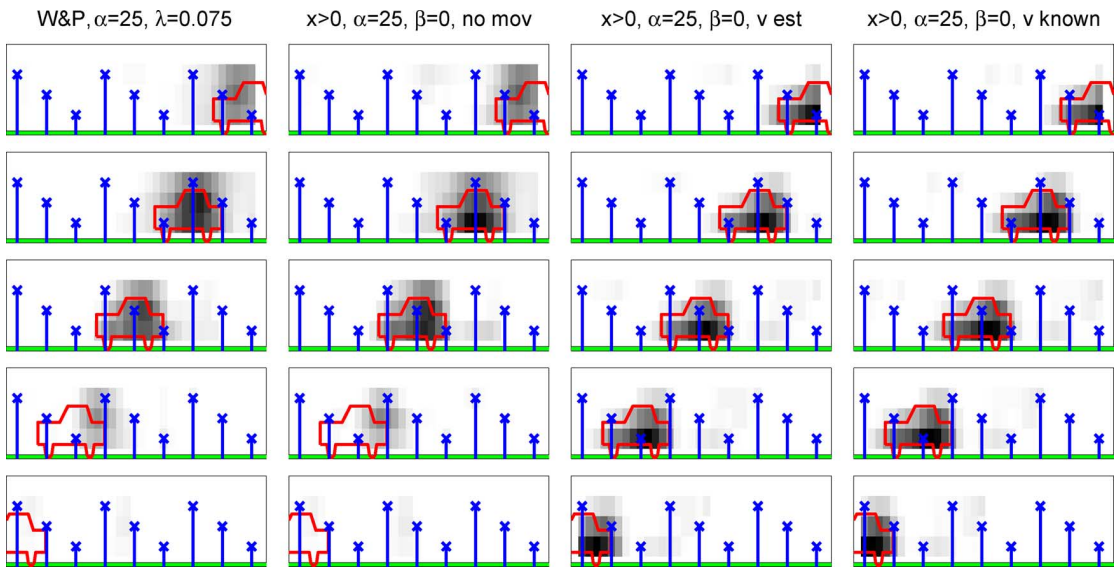

Fig. 11. Experimental results using method (b) from Fig. 8 and no bias term. For the motion constraint, "no mov" ignores the motion and solves each frame separately, "v est" estimates the velocity using (38), and "v known" uses the known true velocity (e.g., by coupling the RTI system with a Doppler sensor).

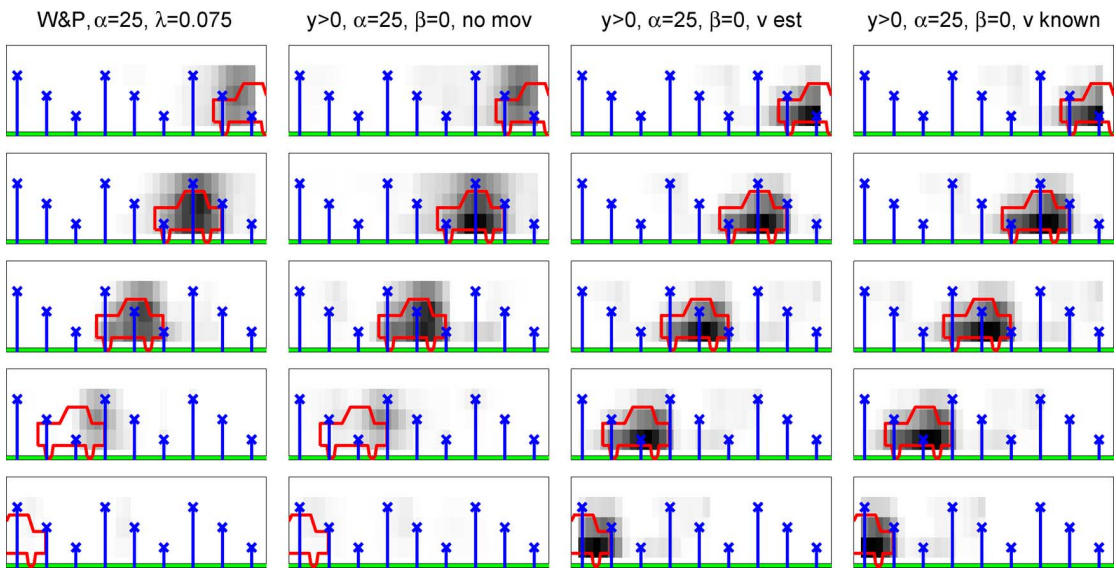

Fig. 12. Experimental results using method (d) from Fig. 8 and no bias term. For the motion constraint, "no mov" ignores the motion and solves each frame separately, "v est" estimates the velocity using (38), and "v known" uses the known true velocity (e.g., by coupling the RTI system with a Doppler sensor).

### B. Subjective Image Quality

This subsection presents a qualitative analysis of Test 1, with the mustang. In most cases, the weight model was the Line model of (4) and (7), though subplots marked "W&P" make use of the NeSh model favored in Wilson and Patwari's work [4]. The method of dealing with negative observations can be to truncate the solution ($\hat{x} \geq 0$), truncate the raw observations ($y \geq 0$), or use the proposed iterative method ("iter"). We considered processing each frame separately ("no mov"), using a movement constraint from (38) with the velocity estimated ("v est"), or using the true velocity ("v known"). These choices and the values of $\alpha$, $\beta$, and $\lambda$ are indicated in each subplot's title.

Results using the NeSh model are included as the first column in Figs. 11 to 13, with analogous results using our preferred Line weight model as the second column in each figure. Due to differences in units between models, the resulting $\mathbf{W}$ matrices were scaled to have comparable magnitude so that the effects of a given value of $\alpha$ would be the same for both models, specifically regarding the term $\mathbf{W}^T\mathbf{W} + \alpha\mathbf{Q}$ in the solution procedure. The final images were also rescaled by dividing the voxel values in any given column of subplots by the maximum of all voxel values in that column. This enables a visual comparison of the results, since otherwise (again, due to differences in units), the images would be scaled differently. The parameter $\lambda$ for [4] was visually optimized to make the estimate match the true scene as well as possible. Changing the weight model leads to a modest visual improvement; and the next subsection will show that there is a corresponding numerical gain.

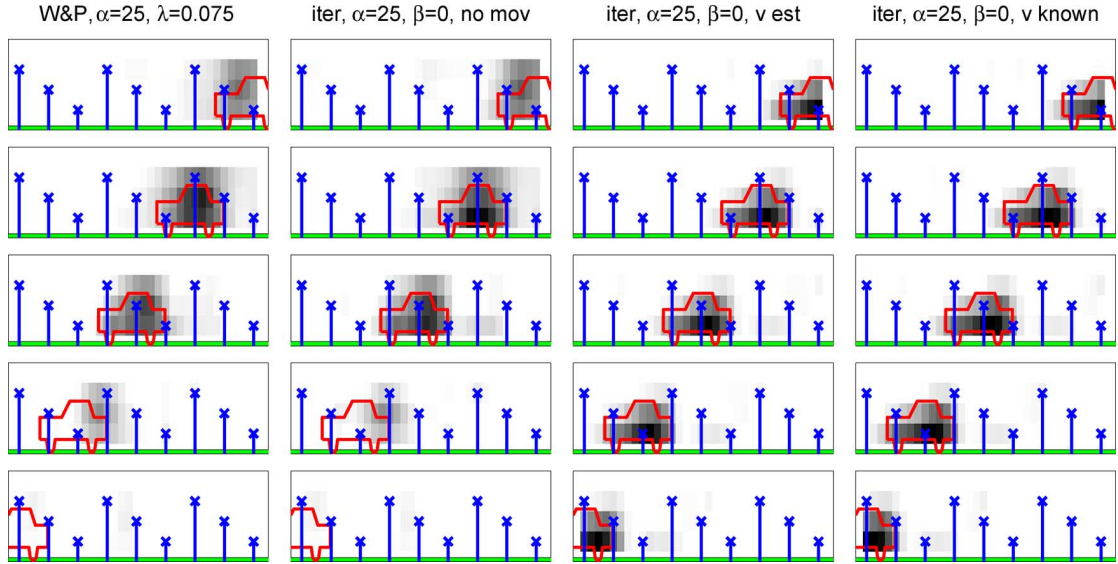

Fig. 13. Experimental results using our proposed iterative method for (c) from Fig. 8 and no bias term. For the motion constraint, "no mov" ignores the motion and solves each frame separately, "v est" estimates the velocity using (38), and "v known" uses the known true velocity (e.g., by coupling the RTI system with a Doppler sensor).

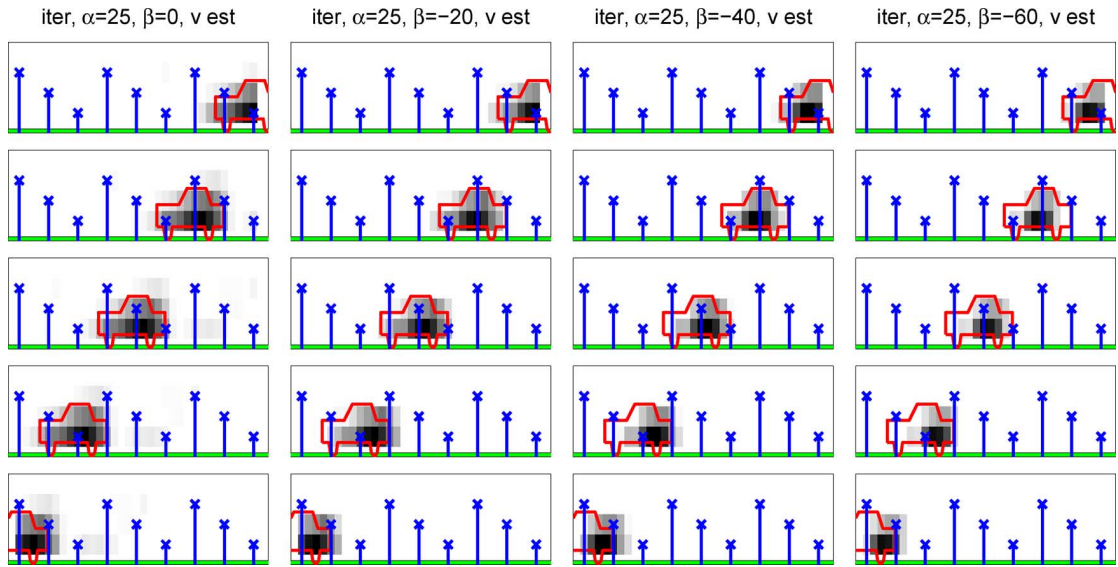

Fig. 14. Experimental results using our iterative method (c) from Fig. 8 with several different values for the bias term.

The biggest factor in image quality is the vehicle motion constraint. This is shown by comparing the third column in each of Figs. 11 to 13 to the second columns. If true velocity is known (fourth column), the results are slightly better than when the velocity must be estimated; this knowledge could be obtained by coupling the RTI testbed with a Doppler sensor. However, even using the estimated velocity provides a vast improvement, effectively increasing the amount of data that can be used to estimate the outline of the vehicle.

To deal with negative observations, Figs. 11 to 13 use methods (b), (d), and (c) from Fig. 8, respectively. Fig. 13 contains similar values for the object voxels as Figs. 11 and 12, but the noise voxels have been largely eliminated.

Fig. 14 shows the results of using various levels of bias. The first column sets the bias to zero (i.e., what is typically done in the literature), whereas a bias value of $\beta = -20$ further suppresses the noise with minimal impact to the quality of the estimate of the voxels within the vehicle. As with the original parameters $\alpha$ and $\lambda$, the value of $\beta$ may need to be re-tuned for other data sets.

Combining the vehicle motion constraint, the iterative solution method to deal with negative data, and the use of the bias term, the final image quality is greatly improved. The baseline case is columns 1 or 2 from Fig. 11 and our recommended implementation is column 2 from Fig. 14.

### C. ATR Performance

Table II shows the results of performing ATR on the estimated images. First the image was quantized to a binary mask; voxels with $\hat{x}_n > 0.275$ dB/m of attenuation were declared to be "occupied," and this threshold was chosen to optimize ATR performance for this data set. A binary true occupancy model was made for each vehicle as well, using the measured outline of the vehicle. Finally, the estimated binary occupancy was compared to each model, and the model with the highest number of voxels in agreement was chosen. This is not necessarily an optimal ATR algorithm; however, the truth data we had available was limited to the shape of each vehicle, and the actual true values of $x_n$ (rather than a binary approximation) could not readily be determined. Table II shows that the Line model for $\mathbf{W}$ performs slightly better than the NeSh model, and the use of the motion constraint improves the results. Note that for Test 8, the last 2 methods had unexpected errors; upon inquiry, we saw many outliers in the data set, which may have been caused by the fact that in this test, the reflectors were removed. Also note that the system always had trouble with the small electric car (Test 3). Due to the variability of materials it contained (a mixture of plastic and metal) as well as its small size, its estimated outline was usually irregular, making it look more like the mustang. This was true regardless of the method of processing.

TABLE II
ATR RESULTS, WITH ERRORS IN CAPITALS. THE "METHOD" FIRST ITEM DENOTES THE NESH OR LINE MODEL FOR $\mathbf{W}$; THE SECOND ITEM DENOTES THE METHOD FOR DEALING WITH NEGATIVE OBSERVATIONS (TRUNCATE $\hat{x}$, TRUNCATE $y$, OR THE ITERATIVE METHOD); AND THE LAST ITEM IS THE NUMBER OF FRAMES USED TO FORM THE IMAGE BEFORE DETECTION [1 OR ALL (A)]

| Test | 1 | 2 | 3 | 4 | 5 | 6 | 7 | 8 | 9 | score |
|---|---|---|---|---|---|---|---|---|---|---|
| method | | | | | | | | | | |
| N/$\hat{x}$/1 | m | v | M | b | V | m | m | m | m | 7/9 |
| L/$\hat{x}$/1 | m | v | M | b | m | m | m | m | V | 7/9 |
| N/$y$/1 | m | v | V | b | V | m | V | m | m | 6/9 |
| L/$y$/1 | m | v | M | b | m | m | m | m | V | 7/9 |
| N/iter/1 | m | v | M | b | V | m | m | m | m | 7/9 |
| L/iter/1 | m | v | M | b | m | m | m | m | m | 8/9 |
| N/$\hat{x}$/A | m | v | M | b | m | m | m | m | m | 8/9 |
| L/$\hat{x}$/A | m | v | M | b | m | m | m | m | m | 8/9 |
| N/$y$/A | m | v | M | b | m | m | m | m | m | 8/9 |
| L/$y$/A | m | v | M | b | m | m | m | m | m | 8/9 |
| N/iter/A | m | v | M | b | m | m | m | E | m | 7/9 |
| L/iter/A | m | v | M | b | m | m | m | E | m | 7/9 |

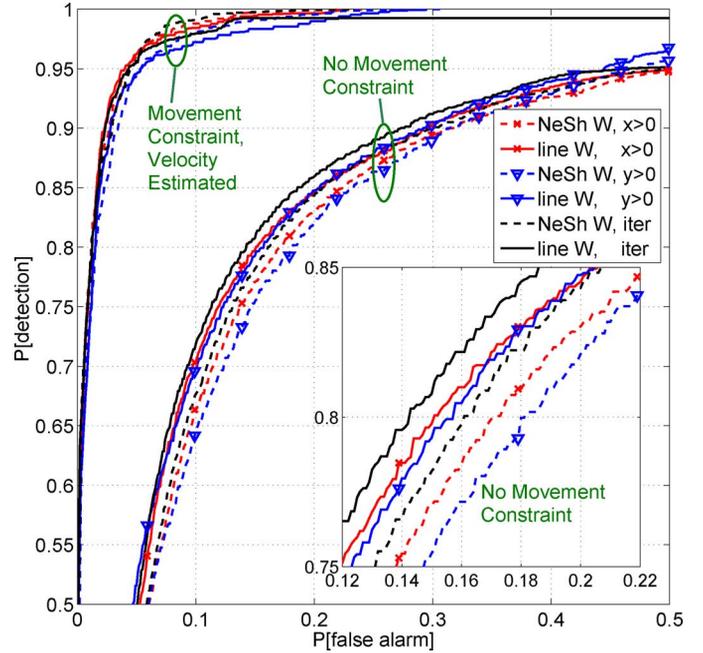

Fig. 15. ROC curves averaged over all the tests with the mustang (tests 1 and 5–9).

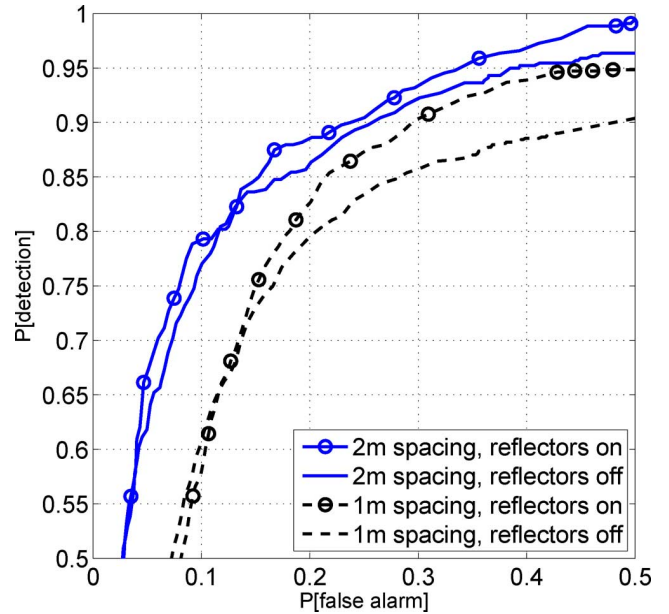

Fig. 16. ROC curves for Tests 6–9 individually, which contrast having the metal reflectors on and off. All curves used the Line model for $\mathbf{W}$, the iterative solving method, and no movement constraint.

Aside from these few errors, this system does appear to be capable of ATR, even though relatively few nodes were used. For a larger system with 2-3 times as many nodes, it may be possible to discriminate between comparably-sized vehicles, though further tests are needed to confirm this.

### D. ROC Curve Analysis

Figs. 15 to 18 show ROC curves [37] for various tests and processing methods. Within the context of the ROC curve, the hypotheses are H0: a given voxel is empty ($x_n = 0$) and H1: a given voxel is occupied ($x_n > 0$). Assuming Gaussian observation noise in (1) and processing of the form of (32), the optimal occupancy detector is a threshold detector of the form

$$\hat{x}_n(\mathbf{y}) \underset{H_0}{\overset{H_1}{\gtrless}} \gamma \qquad (39)$$

where the threshold $\gamma$ is varied over the positive reals to create the ROC curve. The resulting horizontal axis ($P_f$) is the probability that an empty voxel is erroneously declared occupied, and the vertical axis ($P_d$) is the probability that an occupied voxel is correctly declared occupied. The results are averaged over all available frames and all voxels within each frame. In Figs. 15, 17 and 18, averaging over different tests involved averaging all $P_d$ values for a given $P_f$ value. In Fig. 18, the step size $\mu$ was set to slightly lower than the minimum value that caused instability in the updates.

The results in Fig. 15 quantify the performance from the mustang test shown in Figs. 11 to 13. From Fig. 16, the use of the metal reflectors does improve performance slightly by mitigating multipath. From Fig. 17, depending on the operating point, $\beta \in [-30, -10]$ appears to be a good choice, reinforcing Fig. 14. From Fig. 18, PGM ranges from the performance of truncating $\hat{\mathbf{x}} \geq 0$ to the performance of the proposed iterative method depending on the number of iterations, with about 50 iterations needed for convergence.

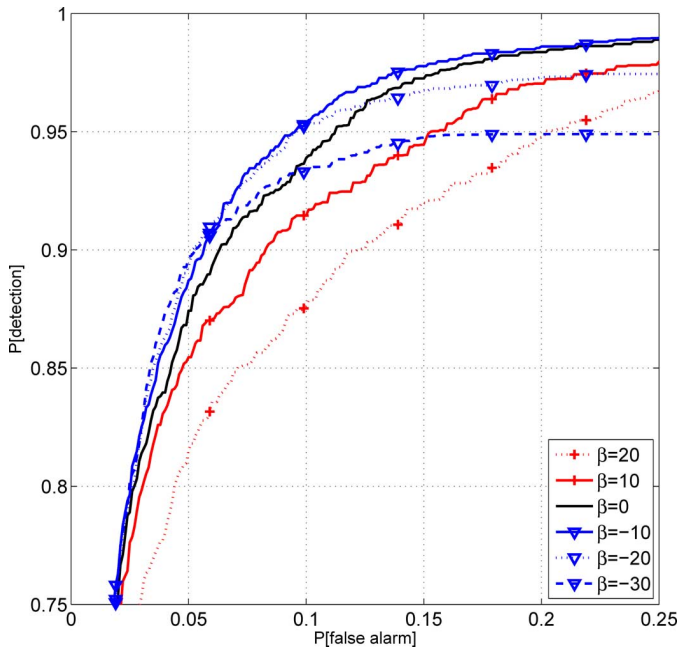

Fig. 17. ROC curves showing the effect of the bias term ($\beta$), averaged over all 9 tests. All curves used the Line model for $\mathbf{W}$, the iterative solving method, and the movement constraint with the velocity estimated.

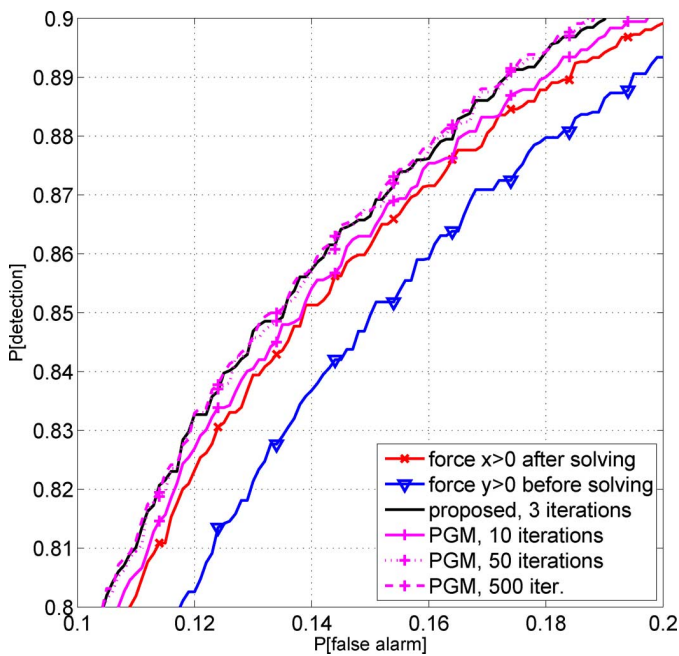

Fig. 18. ROC curves for methods of mitigating negative data, averaged over all 9 tests. All curves used the Line model for $\mathbf{W}$, a step size of $\mu = 0.001$ (if applicable), and no movement constraint.

## VI. CONCLUSION

RTI is a relatively new method for passive all-weather all-illumination localization. This paper proposed a variety of RTI modeling and algorithmic improvements for roadside surveillance, and demonstrated them on a measured data set. The improvements include the use of a more physically motivated weight matrix, a method for mitigating negative data due to noisy observations, and a method for combining the frames of a moving vehicle into a single image. The proposed approaches are used to show improvement in imaging, which would improve human-in-the-loop target recognition; as well as computer-based automatic target recognition.

ACKNOWLEDGMENT

The authors wish to thank Midn E. L. Puzo, Midn R. P. Brenner, and Midn R. P. S. Inglis for assisting with the testbed construction and data collection.

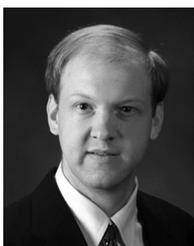

**Christopher R. Anderson** (S'96–M'06–SM'11) joined the United States Naval Academy from Virginia Tech as an Assistant Professor in 2007. In 2013, he was promoted to Associate Professor of Electrical Engineering. He is the founder and current director of the *Wireless Measurements Group*, a focused research group that specializes in spectrum, propagation, and field strength measurements in diverse environments and at frequencies ranging from 300 MHz to over 20 GHz.

Anderson's current research interests include radiowave propagation measurements and modeling, embedded software-defined radios, dynamic spectrum sharing, and ultra wideband communications. He is a Senior Member of IEEE, has authored or co-authored over 30 refereed publications, and his research has been funded by the National Science Foundation, the Office of Naval Research, NASA, and the Federal Railroad Administration.

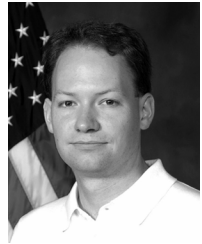

**Richard K. Martin** received dual B.S. degrees (summa cum laude) in physics and electrical engineering from the University of Maryland, College Park, in 1999 and the M.S. and Ph.D. degrees in electrical engineering from Cornell University, Ithaca, NY, in 2001 and 2004, respectively. Since August 2004, he has been with the Department of Electrical and Computer Engineering, Air Force Institute of Technology (AFIT), Dayton, OH, where he is an Associate Professor. He is the author of 29 journal papers and 56 conference papers, and he holds five patents. His research interests include radio tomographic imaging; navigation and source localization; and laser radar. Dr. Martin has been elected Electrical and Computer Engineering Instructor of the Quarter three times and HKN Instructor of the Year twice by the AFIT students. He is currently serving as an Associate Editor for IEEE SIGNAL PROCESSING LETTERS and a Guest Editor for The IEEE JOURNAL OF SELECTED TOPICS IN SIGNAL PROCESSING.

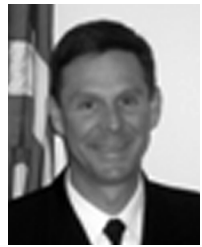

**T. Owens Walker** is a Commander in the United States Navy and an Assistant Professor in the Electrical and Computer Engineering Department of the United States Naval Academy in Annapolis, Maryland. He received his B.S. in Electrical Engineering from Cornell University in 1987 and both his M.S. and Ph.D. in Electrical Engineer degree from the Naval Postgraduate School in 1995 and 2009, respectively. His research interests include wireless sensor networks, mobile ad hoc networks, and wireless medium access. He is a member of the IEEE and Eta Kappa Nu.

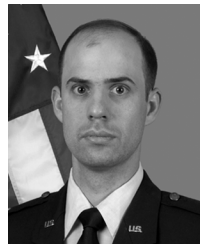

**Ryan W. Thomas** is an Adjunct Professor of Computer Engineering in the Department of Electrical and Computer Engineering, Air Force Institute of Technology, Wright-Patterson AFB, OH. He received his Ph.D. in Computer Engineering from Virginia Polytechnic Institute and State University, Blacksburg, VA in 2007; M.S. in Computer Engineering from the Air Force Institute of Technology, Wright-Patterson AFB, OH in 2001; and his B.S. from Harvey Mudd College in Claremont, CA in 1999. He previously worked at the Air Force Research Laboratory, Sensors Directorate as a digital antenna array engineer. Maj Thomas's research focuses on the design, architecture and evaluation of cognitive networks, cognitive radios and software defined radios.